\pdfoutput=1

\documentclass[11pt]{llncs}

\makeatletter
\def\@seccntformat#1{\@ifundefined{#1@cntformat}%
   {\csname the#1\endcsname\quad}  
   {\csname #1@cntformat\endcsname}
}
\let\oldappendix\appendix 
\renewcommand\appendix{%
    \oldappendix
    \newcommand{\section@cntformat}{\appendixname~\thesection\quad}

}
\makeatother

\usepackage{cite}
\usepackage{makeidx}  
\usepackage{graphicx}
\usepackage{epstopdf}
\usepackage{color}
\usepackage{algorithm}
\usepackage{algorithmic}
\usepackage{multirow}
\usepackage{hhline}
\usepackage{mdframed}
\usepackage{url}
\usepackage{soul}

\newcommand{\myspaceL}{-15pt}
\newcommand{\myspaceM}{-7pt}
\newcommand{\myspaceS}{-1pt}

\begin{document}


\mainmatter  

\title{Ontology-Based Quality Evaluation of \\ Value Generalization Hierarchies \\ for Data Anonymization}

\author{Vanessa Ayala-Rivera\inst{1} \and Patrick McDonagh\inst{2} \and Thomas Cerqueus\inst{1} \and \\ Liam Murphy\inst{1}}

\authorrunning{Vanessa Ayala-Rivera et al.}

\urldef{\mailucdconnect}\path|vanessa.ayala-rivera@ucdconnect.ie,|
\urldef{\mailucd}\path|{thomas.cerqueus,liam.murphy}@ucd.ie,|
\urldef{\maildcu}\path|patrick.mcdonagh@dcu.ie|

\institute{Lero@UCD, School of Computer Science and Informatics, \\ University College Dublin \\
\mailucdconnect\\
\mailucd\\
\and
Lero@DCU, School of Electronic Engineering, Dublin City University \\
\maildcu}

\toctitle{Lecture Notes in Computer Science}
\tocauthor{Authors' Instructions}
\maketitle

\begin{abstract}
In privacy-preserving data publishing, approaches using Va\-lue Generalization Hierarchies (VGHs) form an important class of anonymization algorithms. VGHs play a key role in the utility of published datasets as they dictate how the anonymization of the data occurs. For categorical attributes, it is imperative to preserve the semantics of the original data in order to achieve a higher utility. Despite this, semantics have not being formally considered in the specification of VGHs. Moreover, there are no methods that allow the users to assess the quality of their VGH. In this paper, we propose a measurement scheme, based on ontologies, to quantitatively evaluate the quality of VGHs, in terms of semantic consistency and taxonomic organization, with the aim of producing higher-quality anonymizations. We demonstrate, through a case study, how our evaluation scheme can be used to compare the quality of multiple VGHs and can help to identify faulty VGHs. 
\end{abstract}
\vspace{\myspaceL{}}
\vspace{\myspaceL{}}

\section{Introduction}
\label{sec:Introduction}
\vspace{\myspaceS{}}

Data publishing is an essential element of scientific and societal research. By exploiting data, researchers can create innovative solutions and improved services. However, this data often contains sensitive information about individuals, whose personal data needs to be protected from disclosure. 
Privacy-Preserving Data Publishing (PPDP) develops methods of anonymization for releasing this data without compromising the confidentiality of individuals, while trying to retain the utility of the data. 

A common mechanism to anonymize data is \textit{generalization}. This consists in replacing a specific value with a broader, more general value (e.g., replacing \textit{flu} with \textit{respiratory disease}) with the objective of making the original value more difficult to distinguish. 
\textit{Full-domain generalization} is one of the most known and widely used generalization schemes \cite{LeFevre2005, Machanavajjhala2007, Li2007, Samarati2001, Sweeney2002}.
Under this scheme, all values in an attribute are generalized to their respective ancestor values at the same (higher) level of a hierarchy. This hierarchy, commonly known as \textit{Value Generalization Hierarchy} (VGH) \cite{Samarati2001}, contains a set of terms related to an attribute within a specific domain. The leaf nodes correspond to the original values of a dataset and the ancestor nodes correspond to the candidate values used for the generalizations. More general terms are located at higher levels in the VGH and more specialized terms are lower in the VGH. 

For categorical attributes, a generalization should ideally correspond to a ``less specific but semantically consistent value'' \cite{Sweeney2002}. In spite of this objective, most of anonymization methods do not usually consider the semantics of the terms \cite{Martinez2010}. Some generalization methods rely on the assumption that VGHs are well-specified by preserving the proper semantics in the VGH specification. In this context, it has been discussed in the literature that VGHs play an important role in the quality of the anonymized data \cite{Campan2011, Nergiz2007}. It has also been argued that a ``good'' VGH may improve the utility of the anonymized data \cite{Campan2011}. Similarly, a ``bad'' VGH may cause over-generalization which can potentially reduce data precision \cite{Loh2010a, Nergiz2007}. However, it is unclear as to what a ``good'' or ``bad'' VGH is quantitatively, and how the quality of a VGH can be measured. So far, the responses to these questions have been left to the judgement of the users who define the VGHs. Moreover, these decisions sometimes represent the subjective opinion of a single individual, and thus correspond to just one interpretation of a domain (to which the VGH pertains to). These situations demonstrate how user-defined VGHs can offer a partial and subjective knowledge model of a domain. In our opinion, above problems occur because there are currently no approaches that examine what a ``good'' VGH is, or any other mechanisms that allow the users to assess in a standardized manner the quality of their VGH. This is further exacerbated by the fact that VGHs may be specified without a deep knowledge in the underlying semantics of the domain which the VGH represents.

In this paper, to address the above problems, 
we introduce a method for the evaluation of the quality of VGHs with respect to how well the semantics of the concepts specified in the VGH are maintained throughout the generalization process. The quality of VGHs is measured using semantic similarity metrics applied to the concepts found in the VGH and also using the structural organization of the VGH. To the best of our knowledge, none of the previous works have proposed an approach that applies ontologies and semantics to VGHs to allow users to assess the quality of the VGHs used for anonymization. As a result of measuring the semantic loss of VGHs, the users can improve the specification of their VGHs and prevent applications from using inconsistent, incorrect, or redundant VGHs. Thus, helping to improve the utility of the anonymized data by retaining more meaning of the original concepts. 

The main contributions of this paper are as follows:
\begin{itemize}\itemsep2pt 
\item We propose a new method and a composite score to evaluate the quality of a given VGH based on the semantic properties of the VGH and the information contained in a reference ontology.
\item We analyze and discuss the issues commonly encountered in the specification of VGHs and identify desirable properties in a ``good'' VGH.
\end{itemize}

Section~\ref{sec:MotivAndRelatedWork} discusses the related work and the motivation for the use of semantics and ontologies in anonymization. Section~\ref{sec:RatingVGH} presents our VGH quality assessment method. Section~\ref{sec:PoC} presents our empirical evaluation. 
Section~\ref{sec:Conclusions} presents our conclusions and future work. 

\section{Background and Related Work}
\label{sec:MotivAndRelatedWork}

VGHs for categorical attributes can be manually created by knowledge engineers, domain experts or users, who attempt to preserve the proper semantics in their specification. However, two important aspects with respect to the specification of VGHs remain open and have not been addressed before. First, preserving the underlying semantics of the concepts when defining a VGH and secondly, the existence of a measurement scheme based on standard representations of knowledge that can be used to quantitatively 
evaluate the quality of a VGH.

\textbf{Importance of Semantics in VGHs.}
Preserving the semantics of data is a key requirement when generalizing categorical attributes. Despite its importance, semantics have not been properly or sufficiently considered as part of the anonymization process. Many anonymization methods have ignored this issue by dealing with categorical data in a na\"{\i}ve way, proposing arbitrary suppressions or generalizations that neglect the importance of the semantics of the data \cite{Loh2010a, Martinez2010}. Generalizations may also be carried out using semantically-unaware VGHs (e.g., alphabetically-ordered VGHs), which negatively impact the utility of the anonymized data. For example, consider a list of academic course names. These courses can be generalized into alphabetical ranges (``A-E'', ``F-J'', and so on), according to their first letter. However, such a hierarchy does not make sense, as no information can be acquired from these generalizations (e.g., alphabetical ranges do not provide any useful indication as to which discipline/department each course belongs to). 
This example demonstrates that the quality of the results (and the analysis performed) depends on the VGH definition, thus motivating the importance of using semantically-meaningful VGHs.  However, this is not a trivial task as it can be difficult to identify when these  problematic scenarios may occur. This is because, there are no formal approaches in the current literature to assess the quality of VGHs in the context of anonymization.
Preservation of semantics is a dimension that has shallowly been considered in related works \cite{Martinez2010, Batet2013}. Only recently have researchers investigated this fundamental aspect and integrated this to some degree in the anonymization of categorical attributes \cite{Batet2013, Domingo-Ferrer2013, Han2014, Martinez2010, Martinez2012, FerrerPSD2012}. In some of these works, the use of semantics is often tightly coupled with the proposed algorithms, as they incorporate semantics in the anonymization process itself (execution phase). Our approach incorporates this aspect 
at an earlier stage of anonymization (formalization phase) when the VGH is defined. Moreover, our approach is independent of the methods used for anonymization, as they do not need to be adapted in order to benefit from our VGH evaluation approach. Hence, it is complementary to existing methods, by helping to enhance their effectiveness. If a VGH is semantically coherent, the results can be more meaningful for applications.

\textbf{VGHs: Subjective Knowledge Models.}
Data semantics is defined in \cite{Seth96} as ``the meaning of data and a reflection of the real world''. As users can perceive the real world differently (based on education, cultural background, etc.), there can be more than a single way to represent objects and their relationships. 
For example, in the PPDP area, there have been disagreements about how the VGHs should be specified for a particular domain. In \cite{Fung2005}, Fung et al. did not agree with the groupings specified by Iyengar\cite{Iyengar2002} for the \textit{native-country} attribute. Iyengar grouped the values according to continents, except Americas; whereas Fung et al. followed the grouping according to the World Factbook \cite{worldFactbook}. To avoid this type of discrepancy, VGHs should ideally be created by domain experts who will provide the adequate semantic background for the specification of the VGH. However, this is rarely the case as subject-matter experts are becoming less available, and with the rapid evolution of the domain knowledge field, it is possible that their knowledge may become incomplete or obsolete \cite{Zhou2007}. In previous works, it is commonly assumed that the data publishers are capable of creating VGHs based upon their own knowledge \cite{Loh2010a, Campan2011}. These situations demonstrate how VGHs may be limited in scope, offering a partial and biased view of a domain \cite{Martinez2010}, as they usually represent the understanding of a single individual. To address these problems, we advocate for the use of ontologies as standard knowledge structures to evaluate VGHs in terms of semantics preservation.

\textbf{Ontologies: Standard Knowledge Structures.}
Ontologies are \\structures that model the knowledge of a particular domain. They represent a formal and explicit specification of shared conceptualizations of a domain of interest \cite{Gruber1993}. Since they are usually created from the consensus of multiple experts, they are widely accepted as accurate, impartial representations of a domain. The concepts in ontologies are associated through relationships. The subsumption relationship (\textit{is-a}) constitutes the backbone of an ontology. However, other type of relationships can exist, such as aggregation (\textit{part-of}), synonymy (\textit{synOf}), or other 
application-specific relationships. An example of an ontology can be seen in Appendix~\ref{apx:ontologyFig}. 

For several years, much effort has been devoted to the development of ontologies. Thus, many ontologies are available today \cite{McGuiness2005, DAquin2012} for various domains (e.g., WordNet \cite{Fellbaum1998} for English terms, UMLS \cite{Lindberg1993} for biomedical concepts). WordNet can be used as a lexical ontology for English terms. It contains nouns, verbs, adjectives and adverbs, which are grouped in sets of synonyms, called \textit{synsets}. Synsets represent one underlying lexical concept or a sense of a group of terms (e.g., to refer to the concept expressed by \textit{``a motor vehicle with four wheels usually propelled by an internal combustion engine''}, we could use any of the following terms: \textit{car, auto, automobile, machine} or \textit{motorcar}). 
In WordNet, common semantic relationships connecting noun concepts are referred to as: synonymy (similarity), hypernymy/hyponymy (subsumption) and holonymy/meronymy (aggregation). Appendix~\ref{apx:hypernymSenses} provides an example of the synonyms and hypernyms structure of a noun in WordNet.
Among the semantic relationships, subsumption is the one that provides a potential basis for the construction of a VGH. This is because, when only \textit{is-a} relationships are considered, an ontology becomes a totally ordered taxonomy, where one concept is a subclass of another, which reflects the principle of specialization/generalization. From the above, we believe that the use of ontologies (and their inherent semantics) in the evaluation of VGHs plays a crucial role in the production of anonymized data with maximum utility. In our work, we exploit ontologies (e.g., WordNet) to propose a method to measure the quality of VGHs in an objective way. Some works have started to use the taxonomical structure of the ontologies (instead of user-defined VGHs) to guide an anonymization process \cite{Martinez2010, Martinez2012}. However, these algorithms have been developed/adapted to efficiently handle the complexity of the graph model offered by ontologies. Otherwise, the direct application of ontologies would negatively impact the algorithms' performance (i.e., too costly) and become impractical in real-world. Especially, in some of the existing anonymization algorithms (e.g., \cite{Sweeney2002, LeFevre2005}), where ``the generalisation space is exponentially large according to the depth of the hierarchy, the branching factor, the values and the number of attributes to consider'' \cite{Martinez2012}. 
Thus, our goal here is to evaluate VGHs, not the creation of anonymization algorithms based on ontologies.

\textbf{Semantic Similarity in Ontologies.}
Semantic similarity refers to ``the proximity of two concepts within a given ontology'' \cite{Lee2008AMIA}. 
Several approaches have been proposed for calculating the semantic similarity between two terms in a taxonomy \cite{Meng2013, Budanitsky2006, Sanchez2012}. Among these, path-based measures represent a straightforward way of computing similarity by relying on the path length connecting two concepts. The lower the distance between the concepts, the higher their similarity. Wu and Palmer's metric (WuP) \cite{Wup1994} is a well-known path-based measure that considers the path length and the position of the compared concepts in the taxonomy. The concepts located in a higher level within a taxonomy are given a larger weight (as they are considered less similar) than those in a lower level. 
Refer to Appendix~\ref{apx:wupMetric} for an explanation of the WuP metric and an example of its calculation.
Applied to the PPDP context, we use semantic distance (the inverse of semantic similarity) to quantify how much meaning of the VGH concepts is lost due to generalization operations.
The objective is to quantitatively measure the quality of a VGH using semantic similarity metrics and an ontology. 
\vspace{\myspaceM{}}

\section{VGH Quality Assessment}
\label{sec:RatingVGH}
This section presents the proposed approach to assess the quality of a VGH. We describe how to calculate a quality score to identify whether the generalization relationships in the VGH have been specified with the intent of preserving the semantics of the concepts in the VGH. Thus, the VGHs can be enhanced to potentially improve the data utility in terms of meaning and accuracy.

Applying the concept of semantic distance to the anonymization context, we propose a quality score, called \textit{Generalization Semantic Loss (GSL)}. GSL quantifies how much information is lost (in terms of semantics) when a value in a leaf node (original value) is replaced with a broader value in an ancestor node as a result of generalization using a VGH. From the semantic loss perspective, lower values of GSL are desirable. GSL is measured from leaves to ancestors, following the full-domain generalization process, and considering only the initial and the final state of the data (not the intermediary generalizations performed to achieve the privacy requirement). 
The GSL score for a leaf-ancestor transition~is:
\vspace{\myspaceM{}}
\begin{equation}
\label{eq:TransGSL}
TransGSL(l, a) = 1 - Sim(l,a) 
\vspace{\myspaceS{}}
\end{equation}
where $l$ and $a$ denote the terms at a leaf and ancestor nodes respectively. In this expression, the value of 1 represents the maximum semantic similarity for the WuP metric. If an alternative similarity metric is used, this 1 should be replaced by the maximum value produced by the chosen metric (i.e., the similarity score between a concept and itself). 

Our VGH assessment approach exploits the taxonomical structure of a \textit{reference ontology}; which represents a generalization hierarchy tree with the finest-granularity for a given domain. The reference ontology is used as the source of knowledge, from which the similarity between the terms specified in a VGH will be evaluated. We only use the \textit{is-a} relationships, as the anonymization methods in our scope are ones based on generalization, which is exactly what this type of subclass relationship represents. 
\vspace{\myspaceM{}}
\floatname{algorithm}{Procedure}
\renewcommand{\algorithmicrequire}{\textbf{Input:}}
\renewcommand{\algorithmicensure}{\textbf{Output:}}
\begin{algorithm}
\caption{Computation of $GSL$}
\label{alg:vghAssesment}
\algsetup {
    linenosize=\small
}
\begin{algorithmic}[1]
\REQUIRE Value Generalization Hierarchy $VGH$, reference ontology $O$, syntactic category of the words in the VGH $cat$;
\ENSURE GSL score assigned to the VGH $VghGSL$;
\STATE  $VghGSL$ = 0;
\STATE $h$ = height($VGH$);
\FOR {$i \in [1,h]$}
    \STATE $levelGSL_i$ = 0;
    \STATE $w_i$ = getWeight($i$, $h$);
    \FOR {$l \in$ getLeafNodes($VGH$)}
        \STATE $a$ = getAncestorNodeOfLevel($i$, $l$, $VGH$);
        \STATE $c_l$ = getConceptFromOntology($l$, $O$, $cat$);
        \STATE $H_n$ = getHypernyms($c_l$, $O$);
        \STATE $c_a$ = getConceptFromOntology($a$, $O$, $cat$, $H_n$);
        \STATE \textbf{$transGSL_{la}$ = TransGSL($c_l$, $c_a$);}
        \STATE \textbf{$levelGSL_i$ = max($levelGSL_i$, $transGSL_{la}$);}
    \ENDFOR
    \STATE $VghGSL$ += ($levelGSL_i$ * $w_i$);
\ENDFOR
\RETURN $VghGSL$; 
\end{algorithmic}
\end{algorithm}

\begin{figure}
\centering
\includegraphics[width=4in]{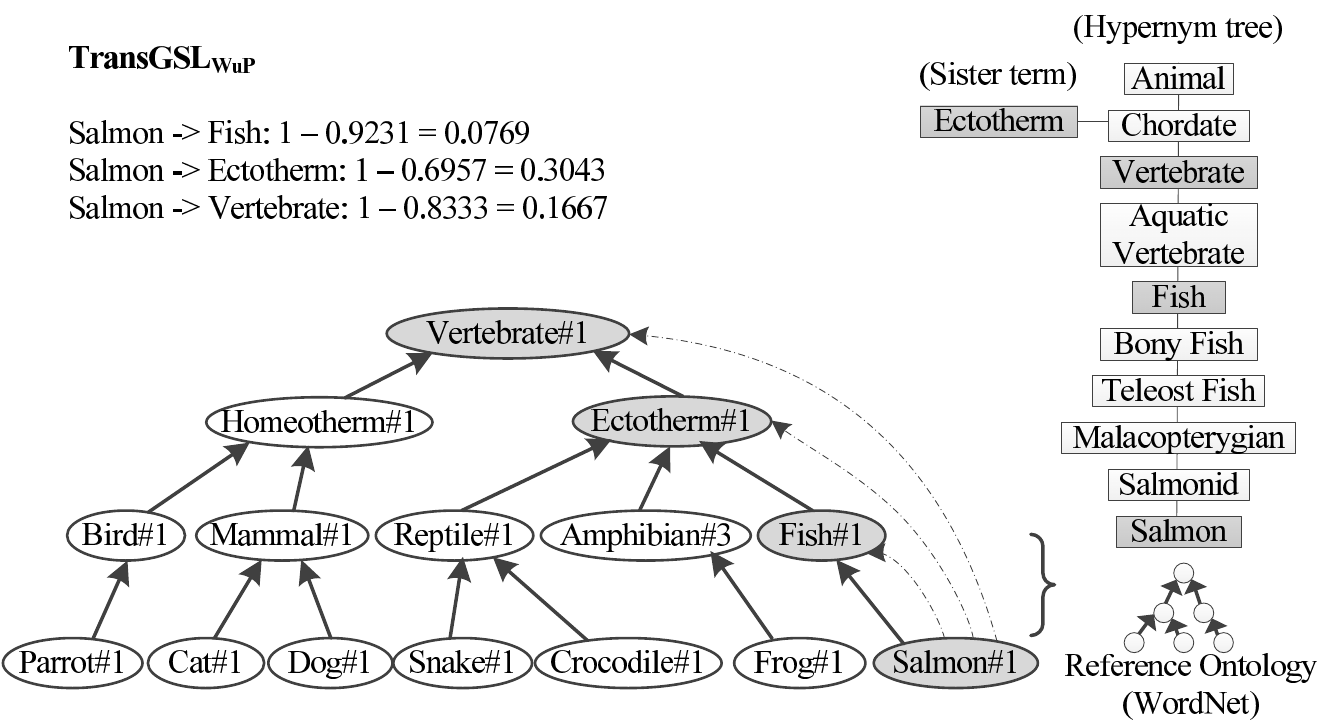}
\vspace{\myspaceM{}}
\caption{Example of TransGSL (leaf-ancestor) calculation in \textit{vertebrates} VGH.}
\label{fig:vghAssessmentSemLoss}
\vspace{\myspaceL{}}
\end{figure}

\vspace{\myspaceM{}}
Procedure \ref{alg:vghAssesment} depicts the process for computing the GSL score for a VGH, termed VghGSL. To aid in the understanding of the quality assessment process, we use a VGH created for a set of vertebrate animals (shown in Figure~\ref{fig:vghAssessmentSemLoss}), \textit{noun} as the syntactic category and WordNet \cite{Fellbaum1998} as the reference ontology. Even though there are limitations to WordNet (e.g., inaccurate or incomplete domain specifications), for the purpose of our experiment, we consider that WordNet represents the \textit{standard} ontology. To calculate semantic similarity, we will use the WuP metric (shown in Appendix \ref{apx:wupMetric}). Compared to other metrics, its simplicity leads to a computationally efficient solution. However, our approach can be applied to other 
similarity metrics.

For each level in the VGH (levels are defined by the height at which the ancestor nodes are positioned in the VGH), the similarity between each leaf and ancestor node needs to be calculated. First, each of the words in the VGH are mapped to a concept (or synset if WordNet is used) in the reference ontology. If the exact word is not found, a synonym is used. 
When multiple senses are available for the same word, the correct sense for the word must be disambiguated. Automatic word-sense disambiguation \cite{Navigli2009} is a broad research field on its own, and is beyond the scope of this paper. In our approach, the senses of the terms at the leaf nodes 
are disambiguated by hand (as the user is involved in the assessment process), consequently the ancestors' senses are derived from the inherited hypernyms associated with the leaf terms.
Appendix~\ref{apx:getSynset} provides a method which demonstrates the retrieval process for a concept.
\vspace{\myspaceL{}}

\begin{figure}
\vspace{\myspaceL{}}  
\centering
\includegraphics[width=4.82in]{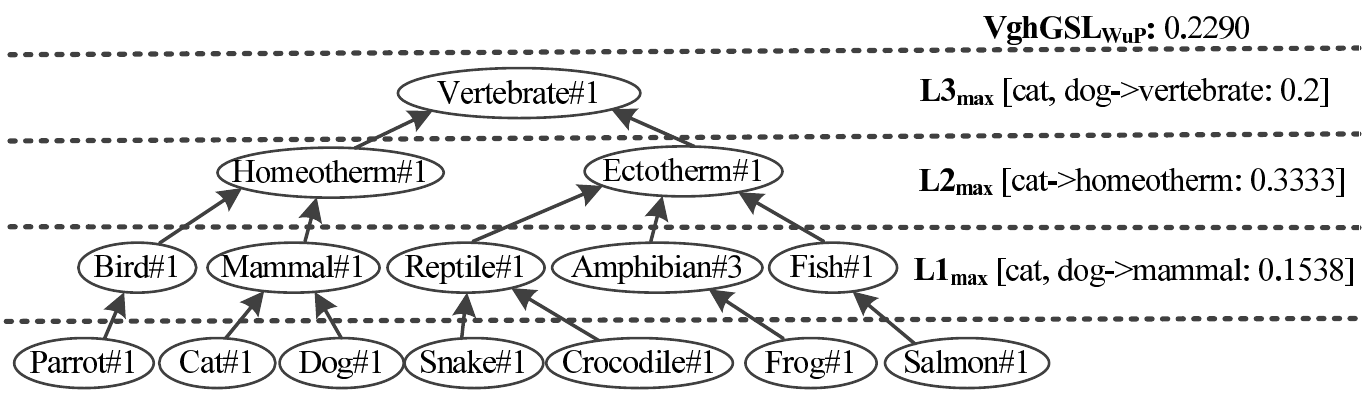}
\vspace{\myspaceL{}}
\caption{Example of LevelGSL calculation in \textit{vertebrates} VGH.}
\label{fig:vghAssessmentLevels}
\vspace{\myspaceM{}}
\end{figure}

Once the correct concepts (and senses) are retrieved from the reference ontology, the GSL scores for leaf-ancestors transitions (TransGSL) are calculated (as given by Equation \ref{eq:TransGSL}). This process is depicted in Figure~\ref{fig:vghAssessmentSemLoss}, which shows an example of how the TransGSL between the leaf node \textit{salmon} (sense\#1), and its corresponding ancestor nodes is calculated using the WuP metric. The semantic similarity is calculated according to WordNet and the associated hypernym tree for \textit{salmon} concept. For example, the semantic similarity between \textit{salmon\#1} and \textit{fish\#1} is 0.9231. Thus, the TransGSL for this transition is 1 - 0.9231 = 0.0769. It can be seen that the TransGSL for the generalization \mbox{\textit{salmon} -$>$ \textit{ectotherm}} is higher than to the other two ancestors. This is because \textit{ectotherm} is not part of the hypernym tree for \textit{salmon} but a sister term (share the same hypernym) of \textit{chordate}. Once the TransGSL scores have been calculated for each leaf-ancestor transition, a representative score for each level is obtained. This is given by: 
\vspace{\myspaceM{}} 
\begin{equation}
LevelGSL(i) = \max_{(l,a)} TransGSL(l, a)
\end{equation}
where $i$ is the index of a level in the VGH, $l$ is a leaf node and $a$ is an ancestor of $l$ in level $i$, and \textit{max} is the maximum score among the TransGSL scores of level $i$. The LevelGSL score is calculated per level, as, in full-domain generalization, the same generalization rules are applied to all values at a particular level of an attribute, such that all values are generalized to their respective ancestor values at the same (higher) level of the VGH \cite{LeFevre2005, Samarati2001, Sweeney2002a}. Moreover, by assessing the semantic loss at each level of the VGH, the users can identify where in the VGH, semantic loss is higher and if required, modify the VGH by referring to the reference ontology. The LevelGSL score can be determined by selecting the score of the transition with maximum loss, or calculating the average loss of all transitions in a level, etc. The choice of the function may depend on the objective of the user. For instance, if we want to avoid the worst cases of semantic loss in the VGH, the maximum score per level can be used to compute the overall (VghGSL) score for the VGH (as shown in Figure~\ref{fig:vghAssessmentLevels}). 
Finally, the LevelGSL scores are multiplied by a weight assigned to each level, and then added up. The VghGSL score is given by:
\vspace{\myspaceM{}}  
\begin{equation}
 VghGSL(VGH) = \sum_{i=1}^{h} w_i \cdot LevelGSL(i)
\end{equation}
where $i$ is the index of a level in the VGH, $w_i$ is the weight associated to Level $i$, and $h$ denotes the height of the VGH. Weights are associated to a level to assign a penalty. The assigned weights have to be specified such that the sum of all weights is equal to 1. 
\begin{table*}
\scriptsize
\caption{TransGSL scores for all levels in the \textit{vertebrates} VGH.}
\label{tbl:vghAssessLevelsTable}
\centering
\begin{tabular}{|l|ccccc|cc|c|}
\hline
\multirow{2}{*}{\textbf{Leaf Nodes}} & \multicolumn{5}{|c}{\textbf{Level 1}} & \multicolumn{ 2}{|c|}{\textbf{Level 2}} & \textbf{Level 3} \\
\hhline{~--------}
 & Bird & Mammal & Reptile & Amphibian & Fish & Homeotherm & Ectotherm & Vertebrate \\
\hline
Parrot & 0.0435 & - & - & - & - & 0.2381 & - & 0.0909 \\
Cat & - & \textbf{0.1538} & - & - & - & \textbf{0.3333} & - & \textbf{0.2} \\
Dog & - & \textbf{0.1538} & - & - & - & 0.1579 & - & \textbf{0.2} \\
Snake & - & - & 0.0833 & - & - & - & 0.2727 & 0.1304 \\
Crocodile & - & - & 0.12 & - & - & - & 0.3043 & 0.1667 \\
Frog & - & - & - & 0.0435 & - & - & 0.2381 & 0.0909 \\
Salmon & - & - & - & - & 0.0769 & - & 0.3043 & 0.1667 \\
\hline
\end{tabular}  
\vspace{\myspaceM{}}
\end{table*}
To explain the VghGSL measure, consider Figure~\ref{fig:vghAssessmentLevels} and Table~\ref{tbl:vghAssessLevelsTable}. The table shows the TransGSL scores calculated for each of the transitions going from the leaf nodes to their corresponding ancestors. Using these scores, the LevelGSL score for each level is calculated according to a function. In this case $max$ (generalization causing the maximum TransGSL), which is shown next to the VGH for each level in Figure~\ref{fig:vghAssessmentLevels}. For example, for Level 1, the maximum value for TransGSL is 0.1538, which is the score corresponding to the generalizations \textit{cat} -$>$ \textit{mammal} and \mbox{\textit{dog} -$>$} \textit{mammal}. In this example, we simplified the calculation of VghGSL by setting all weights to $1/h$ (i.e., $1/3$). The LevelGSL scores are added up (\begin{math}\frac{0.1538}{3} + \frac{0.3333}{3} + \frac{0.2}{3} = 0.2290\end{math}).
\vspace{\myspaceS{}}

\textbf{Weights.} For the computation of VghGSL, we handle two weight variations. The first one is a \textit{constant weight} (i.e., $1/h$), which does not depend on the levels of a VGH, thus, all the levels are penalized in the same manner. This weight is defined with the aim of using the arithmetic mean in the computation of the VghGSL, as it is unknown how many generalizations will be needed to satisfy the privacy requirement. The second variation is a \textit{level-based weight} which depends on the VGH level considered and it is given by: $w_i = \frac{h+1-i}{\sum_{j=1}^{h}j}$, 
where $i$ is the index of a level in the VGH and $h$ denotes the height of the VGH. To better explain how the weights work, consider the case where two VGHs have obtained the same LevelGSL scores, but in different levels. VGH1 has a score of 0.1 and 0.2 in Levels 1 and 2 respectively. VGH2 has the same scores but reversed, this is, 0.2 for Level 1 and 0.1 for Level 2. 
When all the leaf-ancestor transitions in the VGH have the same penalty (using a constant weight, i.e., 1/2), both VGHs obtain the same VghGSL score (0.15). Since we use the average function, the assessment will provide similar scores for correctly (i.e., VGH1) and incorrectly (i.e., VGH2) ordered VGHs. Most of the similarity metrics consider the fact that concepts at the lower levels are more similar than those at the upper levels (e.g., WuP). In order to reintroduce this aspect in our assessment, we penalize the loss of information per level, giving a larger weight to the lower levels, compared to the higher levels.  By using the level-based weight in this scenario (0.666 for Level 1 and 0.333 for Level 2), the VghGSL score is 0.1333 for VGH1 and 0.1666 for VGH2.



\section{Empirical Evaluation}
\label{sec:PoC}
\vspace{\myspaceM{}}
To evaluate our proposed method, we conducted a case study using members from our research group. We pursued two objectives in this experiment: (i) to investigate how VGHs (of the same domain) created by different people are subjective to their interpretation of the domain, and (ii) to demonstrate how our proposed VGH assessment method can be applied to quantitatively measure the quality of the created VGHs. We present the study in two phases. First, we review some of the issues encountered in the specification of a categorical VGH, and second, we show how the VghGSL score can be used to compare in a standard manner, the quality of multiple VGHs. 
Thus, helping to identify which VGH (among a set of VGHs created for a domain) can retain higher utility in the anonymized data by better preserving the semantics of the original values.

For our evaluation, consider the scenario where a veterinary laboratory has been testing a new treatment for animals. The laboratory would like to share their results, 
while protecting the specific details about the animals used in their tests; thus the dataset needs to be anonymized.
%

\textbf{Phase 1: Specification of VGHs. }
To guide the anonymization of the \textit{animal} attribute, we asked two members of our team (postdoctoral researchers who are not experts in the field of knowledge engineering) to create their own VGHs using multiple sources (e.g., dictionaries, Wikipedia, WordNet\footnote{WordNet was used by the subjects only as source of knowledge (e.g., definitions, taxonomies), and not to measure similarity between terms.}) and their own knowledge about the domain. It is worth mentioning that the subjects (i.e., researchers) created the VGHs without pre-computing the semantic loss, or any other information metrics, among the terms in their VGHs. The VGHs created are provided in Appendix \ref{apx:vghfigs} and are denoted as VGH1 and VGH2. The leaf nodes correspond to the original values of \textit{animal} attribute. 
The VGHs created are height-unbalanced (i.e., leaf nodes are at different heights). Since a common pre-condition of full-domain generalization methods is that the VGHs are height-balanced, a typical approach is to replicate the leaf values 
until reaching the same height of the deepest leaf node. 

As discussed in Section \ref{sec:MotivAndRelatedWork}, it is common that data publishers (who are not necessarily domain experts) create a VGH with the aim of anonymizing a dataset. In our experiment, the subjects were not experts in the domain, so they faced some difficulties while defining their VGHs. It was reported that the process of building a VGH from multiple sources was cumbersome, as different taxonomies were available for the same domain. Most of these taxonomies were application-specific, so it was challenging to come up with a final aggregated taxonomy. Another issue in the definition of the VGHs was that the subjects often used adjectives as the terms of the ancestor nodes, which modify or elaborate the meaning of words, rather than representing an \textit{is-a} relationship. This caused the VGHs to have mixed syntactic categories (e.g., nouns and adjectives) in the definition of the ancestor nodes. It has been argued that language semantics are mostly captured by nouns, therefore, most of research focuses on nouns in semantic similarity calculation \cite{Meng2013}. This is the case for WordNet-based similarity metrics. Since these metrics are focused on taxonomic relations, their applicability is restricted to the noun and verb categories.
Moreover, the categories to be measured have to be of the same type (i.e., noun-noun or verb-verb). Therefore, we nominalized the adjectives found in the VGHs mapping them to a related noun, for example \textit{warm-blooded} was mapped to \textit{homeotherm}; similarly \textit{cold-blooded} was mapped to \textit{ectotherm}. Even though the subjects attempted to provide the adequate generalizations in the VGH, in the end, they were uncertain about the quality of their VGHs. Thus, the second phase of our experiment was to compare the quality of the VGHs using our proposed VghGSL measure.

\textbf{Phase 2: Comparing the Quality of VGHs. }
In our implementation, we used WordNet 3.0 and the Java libraries JAWS 1.3 \cite{jaws} and RiTa \cite{rita} to retrieve data from the WordNet database. To calculate the semantic similarity among terms, we used the library JWI \cite{jwi}. 

To compute the VghGSL score, we used the weight variations explained in Section \ref{sec:RatingVGH}. The constant weight to assign no penalty (setting all level weights to $1/h$), and the level-based weight to penalize more the information loss at lower levels (using the $w_i$ equation). 
To compare the VGHs, we first calculated the TransGSL score for all leaf-ancestor transitions and then obtained the LevelGSLs (using the \textit{max} function). Table~\ref{tbl:VGH1VGH2Table} presents the results for each VGH,  
showing the transitions causing the maximum loss per level, and the LevelGSL scores calculated using the constant weight ($1/4$) and the level-based weights (0.4 for Level 1, 0.3 for Level 2, 0.2 for Level~3 and 0.1 for Level 4). 
The VghGSL scores are shown in  the last row of each VGH table.
\begin{table*}
\scriptsize
\vspace{\myspaceM{}}
\caption{VGHs Comparison using Constant and Level-Based Weighted GSL.}
\label{tbl:VGH1VGH2Table}
\centering
\begin{tabular}{|c|l|l|l|}
\hline
\multirow{2}{*}{\textbf{Generalization}} & \multicolumn{3}{|c|}{\textbf{VGH1}} \\
\hhline{~---}
 {} & {\bf Max TransGSL Transition} & {\bf LevelGSL $\cdot 1/h$}  & {\bf LevelGSL $\cdot w_i$} \\
\hline
         L0-$>$L1 & Horse, Giraffe -$>$ Ungulate & 0.0258 & 0.0414  \\
         L0-$>$L2 & Horse, Giraffe, Tiger -$>$ Mammal & 0.0463 & 0.0556  \\
         L0-$>$L3 & Horse, Giraffe, Tiger -$>$ Homeotherm & 0.09 & 0.072  \\
         L0-$>$L4 & Horse, Giraffe, Tiger -$>$ Animal & 0.0833 & 0.0333 \\
\hline
 {\bf VghGSL Score} &  & 0.2454 & 0.2023  \\
\hline
\multicolumn{4}{c}{}\\
\hline
\multirow{2}{*}{\textbf{Generalization}} & \multicolumn{3}{|c|}{\textbf{VGH2}} \\
\hhline{~---}
 {} & {\bf Max TransGSL Transition} & {\bf LevelGSL $\cdot 1/h$} & {\bf LevelGSL $\cdot w_i$} \\
\hline
         L0-$>$L1 & Horse, Giraffe -$>$ Herbivore & 0.09 & 0.1440 \\
         L0-$>$L2 & Horse, Giraffe, Tiger -$>$ Mammal & 0.0463 & 0.0556 \\
         L0-$>$L3 & Horse, Giraffe, Tiger -$>$ Vertebrate & 0.0577 & 0.0462 \\
         L0-$>$L4 & Horse, Giraffe, Tiger -$>$ Animal & 0.0833 & 0.0333\\
\hline
 {\bf VghGSL Score} &  & 0.2773 & 0.2791 \\
\hline
\end{tabular}  
\end{table*}
\begin{figure}[ht]
\vspace{\myspaceM{}}
    \begin{minipage}[b]{0.50\linewidth}
    \centering
    \includegraphics[width=\textwidth]{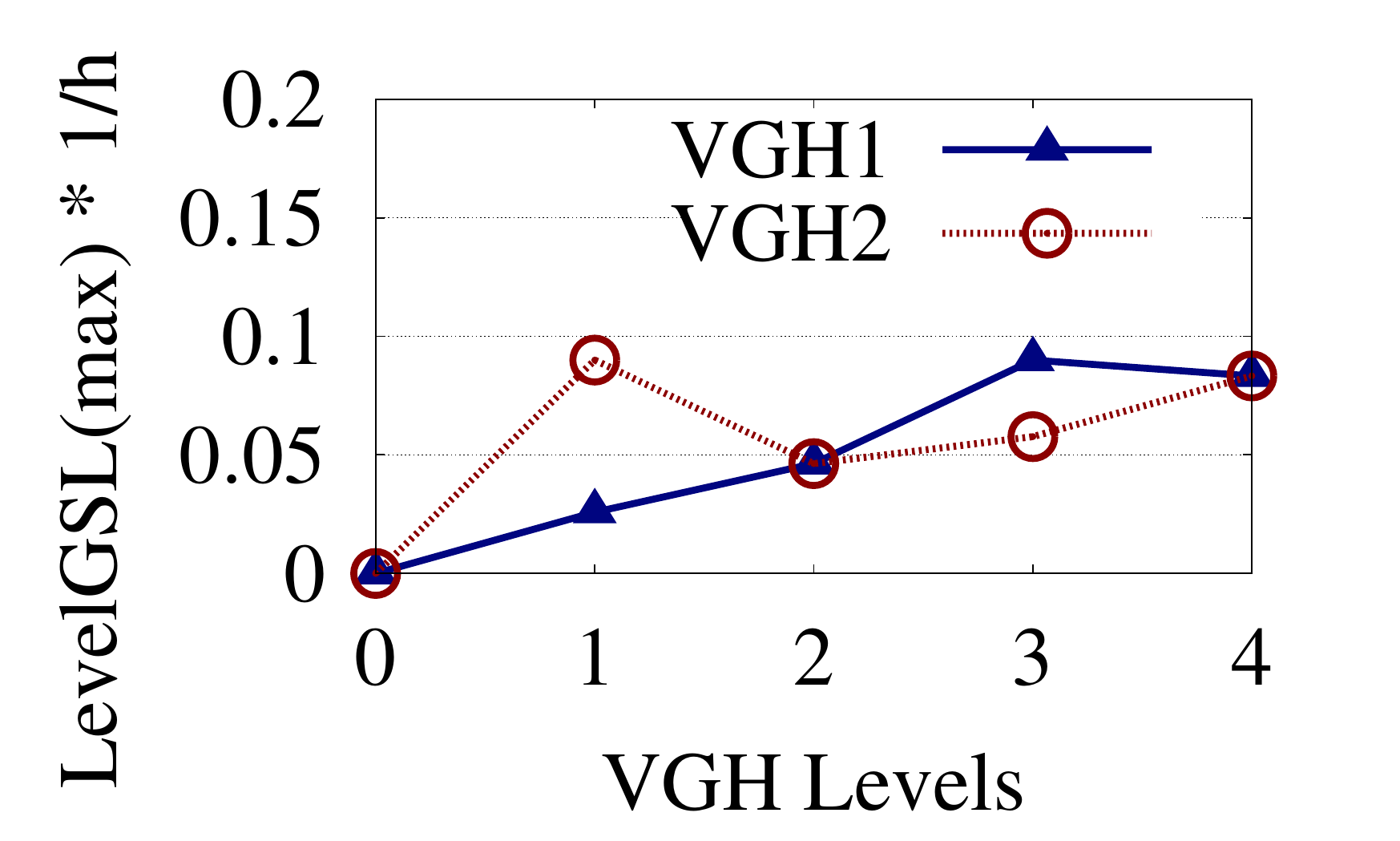}
    \vspace{-20pt}
    \caption{Constant Weight LevelGSLs.}
    \label{fig:VGH1VGH2PlotMaxStCons}
    \vspace{\myspaceL{}}
\end{minipage}
\hspace{0.3cm}
\begin{minipage}[b]{0.50\linewidth}
    \centering
    \includegraphics[width=\textwidth]{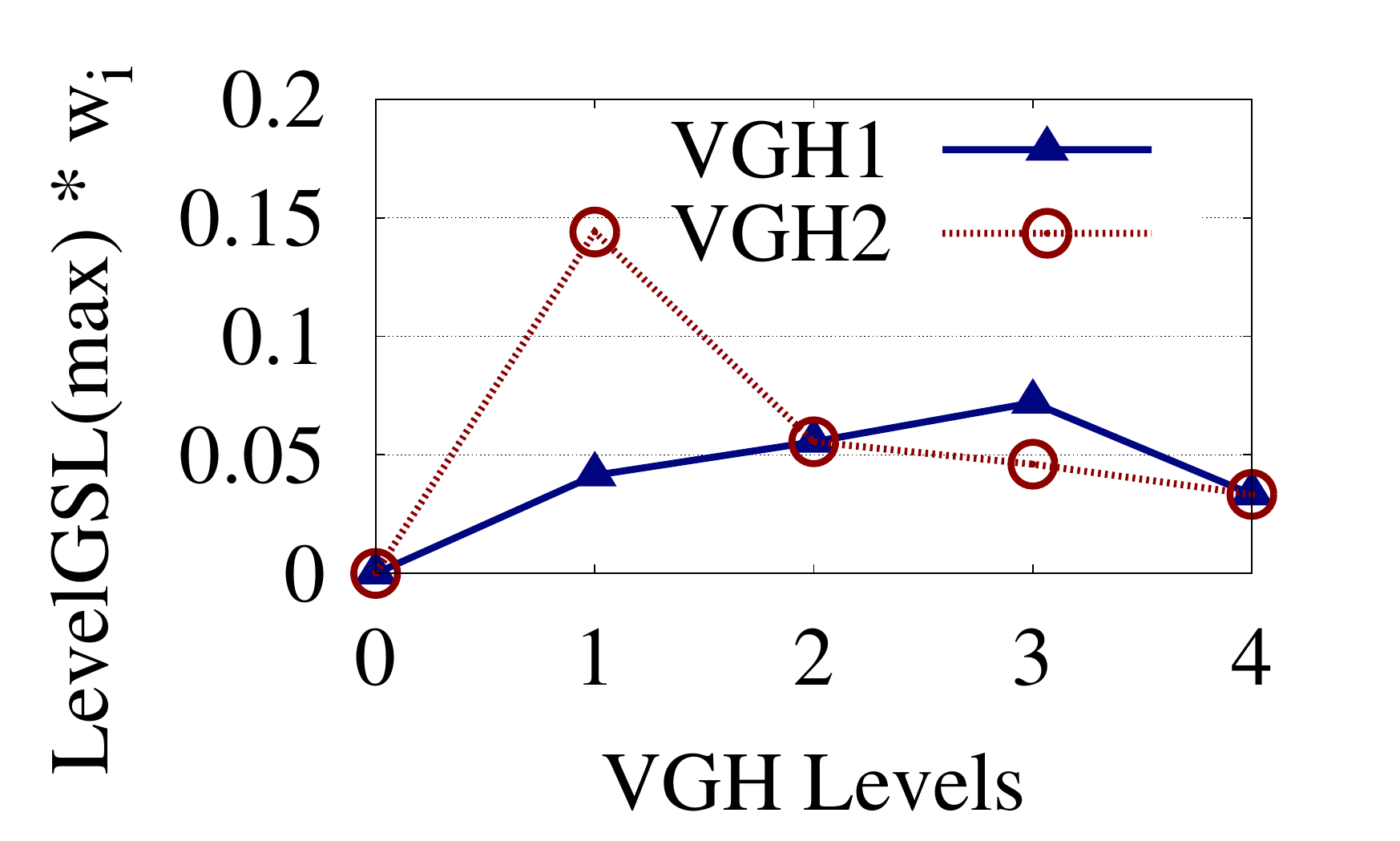}
    \vspace{-20pt}
    \caption{Level-Based Weight LevelGSLs.}
    \label{fig:VGH1VGH2PlotMaxStLevel}
    \vspace{\myspaceL{}}
    \end{minipage}
\end{figure}

From Table~\ref{tbl:VGH1VGH2Table}, it can be deduced that VGH1 is better specified than VGH2. According to the VghGSL scores, VGH1 better preserves the semantics of the original data throughout the generalizations by minimizing the worst cases of semantic loss. However, if we look at the constant weight LevelGSL scores (shown in Figure~\ref{fig:VGH1VGH2PlotMaxStCons}), it can be seen that the scores fluctuate between the VGHs, depending on the number of generalizations required to satisfy the desired privacy degree (e.g., the $k$ value from $k$-anonymity \cite{Samarati2001, Sweeney2002a}). For example, if only one generalization is performed (i.e., ending at Level~1), VGH1 seems to be better than VGH2; however, this situation changes if three generalizations are required (i.e., ending at Level 3). Moreover, the peak observed for VGH2 denotes a poorly-defined generalization, as the score at Level $i$ is higher than the one at Level $i+1$ (i.e., a child concept is less specific than its parent concept). 
Although both VGHs obtained LevelGSL scores of 0.09 (VGH1 at Level~3 and VGH2 at Level~1), these do not represent the same semantic loss in the VGH. Thus, following the idea behind most semantic similarity metrics (i.e., the concepts' meaning is better preserved at the lower levels), we differentiate between the loss at the various levels of the VGH by using the level-based weight($w_i$) LevelGSL. Figure~\ref{fig:VGH1VGH2PlotMaxStLevel} depicts these results, showing that the 0.09 score obtained in lower levels (VGH2 at Level 1) represents a worse case, as losing semantics at lower levels is undesirable (the meaning of the most specific concepts is lost).

In our experiments, we used the \textit{max} function to obtain the LevelGSL, as the aim was to avoid the worst cases of semantic loss. However, other functions can be used. For example, consider the case where most transitions in a VGH have fine-grained definitions (having low semantic loss), except for one branch (transitions forming a path from a leaf to the VGH root). 
Such transitions represent the maximum TransGSL at each level, thus, their scores become the LevelGSLs. In this case, the VGH will be heavily impacted by the high scores in that branch; even when most of the transitions are balanced with a low semantic loss. 
Considering this scenario, a more fair approach would be to use \textit{avg} as the function for LevelGSL selection and complement the results with the \textit{max} and standard deviation per level. 

Finally, in terms of semantic preservation, fine-grained VGHs would be preferable. However, in terms of privacy, this may not be always desirable, as the data may become vulnerable to attacks. Inferences about the data can still happen if the semantic distance between concepts is small enough for the data to be still sensitive (e.g., \textit{crocodile}-$>$\textit{crocodilian}). Ultimately, the users will decide about the specification of their VGHs. 
Our approach will help users to make an informed decision about this by quantitatively assessing VGHs and allowing for comparison between VGHs.

\vspace{\myspaceM{}}
\section{Conclusions And Future Work}
\label{sec:Conclusions}
\vspace{\myspaceM{}}
In this paper, we proposed the use of semantic retention for the evaluation of Value Generalization Hierarchies (VGHs) for categorical attributes. We integrate semantic similarity metrics and the taxonomical structure of ontologies to compute a measure that serves as the quality score for a VGH. This measure quantifies the semantic loss incurred when the original values of a dataset are replaced by broader values due to generalization. Our evaluation shows how our proposed measure can be used to identify VGHs that have not been well-specified, in terms of semantics. Moreover, this measure can be used to compare multiple VGHs in a standard manner and thus help to identify which one better preserves the semantics of the original data. 
Future work involves evaluating the improvements that our VGH assessment approach brings to the utility of the anonymized data in terms of semantics. 
We also intend to explore how to automatically generate semantic-driven VGHs for categorical attributes, 
based on ontologies. 
We also plan to consider other semantic similarity measures for our VGH assessment method.
%
%
\bibliographystyle{abbrv}
\bibliography{PSD2014_paper7_vayala} 

\appendix

\section{Example of a Simple Ontology}
\label{apx:ontologyFig}
\begin{figure}
\centering
\includegraphics{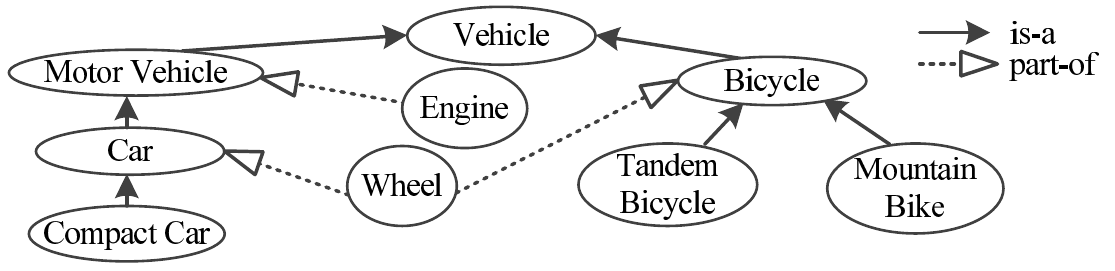}
\vspace{\myspaceL{}}
\caption{An example of a simple ontology for \textit{vehicle}.}
\label{fig:ontologyExample}
\end{figure}

Figure \ref{fig:ontologyExample} shows an example of an ontology having subsumption (\textit{is-a}) and aggregation (\textit{part-of}) relationships.


\section{Hypernym of Senses in WordNet}
\label{apx:hypernymSenses}
This appendix shows an example of how synonyms and hypernyms are structured in WordNet. Figure~\ref{fig:hypernymSenses} provides part of the synonyms and hypernyms for the \textit{bow} noun, showing two different senses: \textit{reverence} and \textit{decoration}.
\begin{figure}
\begin{mdframed}
\setlength{\columnsep}{10pt}
\textbf{Sense 6} \\
\textbf{Bow:} Bending the head or body or knee as a sign of reverence or submission or shame or greeting. \\
$\Rightarrow$ reverence\\
$~~\Rightarrow$ action\\
$~~~~\Rightarrow$ act, deed, human action, human activity\\
$~~~~~~\Rightarrow$ event\\
$~~~~~~~~\Rightarrow$ psychological feature\\
$~~~~~~~~~~\Rightarrow$ abstraction, abstract entity\\
$~~~~~~~~~~~~\Rightarrow$ entity\\
\textbf{Sense 8} \\
\textbf{Bow:} A decorative interlacing of ribbons. \\
$\Rightarrow$ decoration, ornament, ornamentation\\
$~~\Rightarrow$ artifact, artefact\\
$~~~~\Rightarrow$ whole, unit\\
$~~~~~~\Rightarrow$ object, physical object\\
$~~~~~~~~\Rightarrow$ physical entity\\
$~~~~~~~~~~\Rightarrow$ entity
\end{mdframed}
\vspace{\myspaceM{}}
\caption{A hypernym of senses of \textit{bow} in WordNet.}
\label{fig:hypernymSenses}
\end{figure}


\section{The WuPalmer Metric}
\label{apx:wupMetric}
This appendix presents the equation for the WuPalmer measure given by:
\begin{eqnarray*}
Sim_{WuP}(c_{1}, c_{2}) = \frac{2 * N_{3}}{N_{1} + N_{2} + 2 *  N_{3}}
\end{eqnarray*}
where $c_{1}$ and $c_{2}$ are the two concepts for which the semantic similarity is measured, $N_{1}$ and $N_{2}$ denote the number of \textit{is-a} links on the path from $c_{1}$ and $c_{2}$ respectively, to their least common subsumer (LCS), and $N_{3}$ denotes the number of \textit{is-a} links on the path from the LCS to the root of the taxonomy. The score range is (0,1] (1 for identical concepts).
To illustrate the WuP metric, say we want to calculate the similarity between \textit{car} and \textit{compact car} in the ontology shown in Appendix \ref{apx:ontologyFig}. 
The LCS is \textit{car}. Thus, 
following the formula, we obtain \begin{math}Sim_{WuP}(car, compact \hspace{1mm} car) = \frac{2 * 2}{0 + 1 + (2 * 2)} = 0.8\end{math}. 

\section{Retrieving a Concept from WordNet}
\label{apx:getSynset} 
In our work, the senses of the terms at the ancestor nodes are obtained from the inherited hypernyms associated with the leaf terms. To do this, a matching is performed between the hypernyms of a leaf node and each of the leaf node's ancestors. If there is a match, the sense for the matched hypernym is selected. Otherwise, a manual disambiguation is needed. This process is shown below in Procedure 2.
\vspace{\myspaceM{}}
\floatname{algorithm}{Procedure}
\renewcommand{\algorithmicrequire}{\textbf{Input:}}
\renewcommand{\algorithmicensure}{\textbf{Output:}}

\begin{algorithm}
\caption{getConceptFromOntology}
\label{alg:getSynset}
\algsetup {
    linenosize=\small
}
\begin{algorithmic}[1]
\REQUIRE a node in the VGH $n$, reference ontology $O$, syntactic category of the words in the VGH $cat$, inherited hypernyms of the concept in a VGH node $H_n$;
\ENSURE underlying lexical concept from ontology for the VGH node $concept$;
\STATE	$C_n$ = getConceptSetForWord($n$, $O$, $cat$);
\IF {$n$ is a leaf node}	
	\STATE	$s_n$ = getDisambiguatedSense($n$, $C_n$);
\ELSE		
	\IF {$C_n$ is found in $H_n$}
        \STATE $s_n$ = getSense($C_n$, $H_n$);
	\ELSE	
        \STATE $s_n$ = getDisambiguatedSense($n$, $C_n$);
	\ENDIF
\ENDIF	
\STATE	$concept$ = getConcept($s_n$, $C_n$);
\RETURN $concept$;
\end{algorithmic}
\vspace{\myspaceS{}}
\end{algorithm}

\vspace{\myspaceL{}}

\section{VGHs Created for Our Empirical Evaluation}
\label{apx:vghfigs} 
This appendix shows the VGHs created for the \textit{animal} attribute. 
\begin{figure*}
\centering
\includegraphics[width=4.82in]{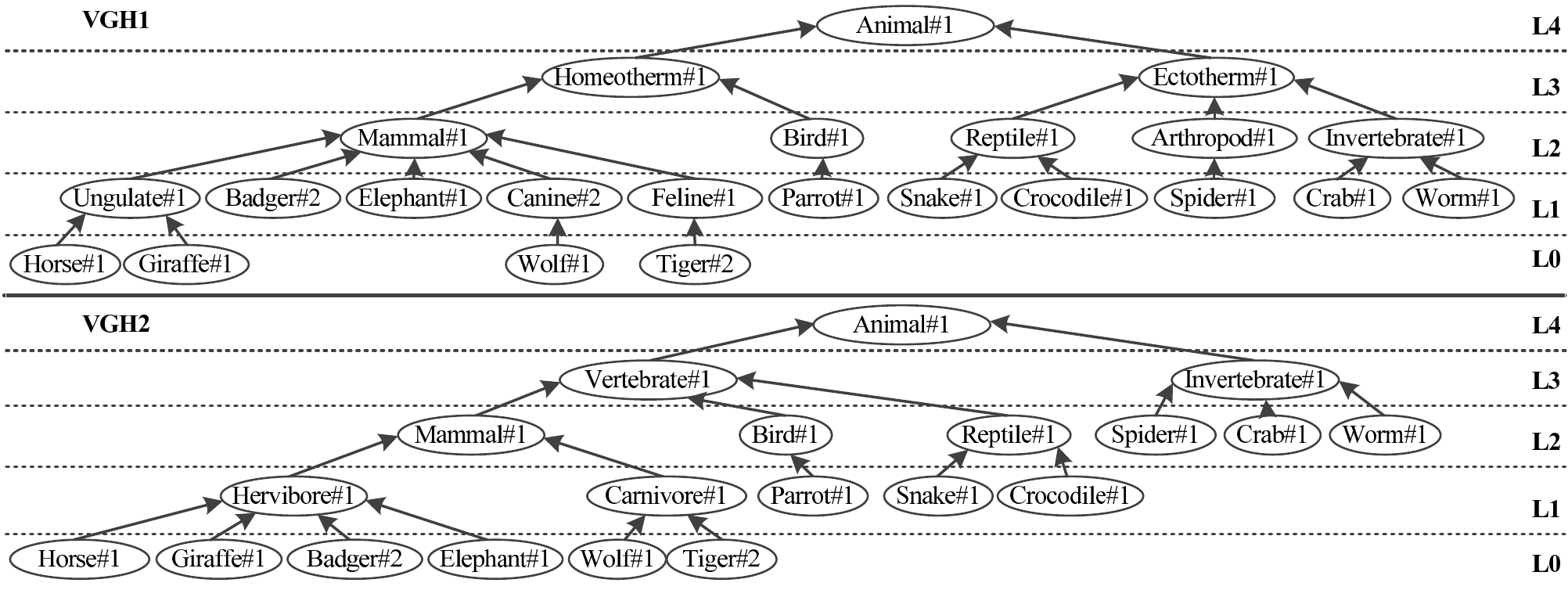}
\vspace{\myspaceL{}}
\caption{The two different VGHs specified in our experiment.}
\label{fig:VGH1VGH2}
\vspace{\myspaceM{}}
\end{figure*}

\vspace{\myspaceL{}}
\section*{Acknowledgments}
Supported, in part, by Science Foundation Ireland grant 10/CE/I1855 and Science Foundation Ireland grant 08/SRC/I1403 FAME SRC (Federated, Autonomic Management of End-to-End Communications Services - Scientific Research Cluster).
\end{document}